# Transition metal dichalcogenide metamaterials with atomic precision

Battulga Munkhbat[1], Andrew B. Yankovich[1], Ruggero Verre[1], Eva Olsson[1], and Timur O. Shegai[1,*]

[1]Department of Physics, Chalmers University of Technology, 412 96, Gothenburg, Sweden

*Correspondence to: timurs@chalmers.se

## ABSTRACT

The ability to extract materials just a few atoms thick has led to discovery of graphene, monolayer transition metal dichalcogenides (TMDs), and other important two-dimensional materials. The next step in promoting understanding and utility of the flatland physics beyond the state-of-the-art is to study one-dimensional edges of such two-dimensional materials as well as to control the edge-plane ratio. Edges typically exhibit properties that are unique and distinctly different from those of planes and bulk. Thus, controlling them allows to design principally new materials with synthetic edge-plane-bulk characteristics, that is, TMD metamaterials. However, the enabling technology to study such metamaterials experimentally in a precise and systematic way has not yet been developed. Here we report a facile and controllable anisotropic wet etching method that allows scalable fabrication of TMD metamaterials with atomic precision. We show that TMDs can be etched along certain crystallographic axes, such that the obtained edges are atomically sharp and exclusively zigzag-terminated. This results in hexagonal nanostructures of predefined order and complexity, including few nanometer thin nanoribbons and nanojunctions. The method thus enables future studies of a broad range of TMD metamaterials with tailored functionality through atomically precise control of the structure.

TMDs are used in numerous research and technology disciplines, as motivated by their excellent optical (*1, 2*), mechanical (*3*), electronic (*4, 5*), nanophotonic (*6*), and catalytic (*7-10*) properties. Their ability to be exfoliated down to monolayer thickness and form van der Waals (vdW) heterostructures (*11*), which in turn give rise to intriguing Moiré physics (*12*), opens possibilities for designing new materials with tailored properties and novel functionalities. The most popular TMDs, such as $MoS_2$ and $WS_2$, are semiconducting in their 2H phase and significant effort is thus focused on exploring their excitonic properties.

The capability to control the edges of such vdW materials is interesting for numerous applications. For example, density functional theory (DFT) predicts that zigzag-terminated edges are metallic and ferromagnetic, while armchair edges are semiconducting and nonmagnetic (*13-15*). The metallic edges also exhibit high electro- and photo-catalytic activity (*16*), notably as non-precious analogs of platinum for the hydrogen evolution reaction (*7-9*). Due to symmetry reduction at the edges, they also show potential for nonlinear optics applications, such as enhanced second harmonic generation (*17*). To utilize edges for catalysis and sensing applications, it is important to control their composition, termination, and edge-plane ratio. Many different methods, including chemical (*18, 19*), surface engineering (*20*), thermal annealing (*21, 22*), plasma treatment (*23*), and other emerging techniques (*24*), have been employed recently to engineer edges of various vdW materials and their monolayers. However, the possibility to deterministically and reproducibly control the edges of TMD materials and edge-plane ratio over large scales remains limited. Such an ability would open multiple possibilities for tailoring material properties through precise control of the structure and would in general allow to formulate a vision of TMD metamaterials.

Here, we introduce a method that allows accurate control over the edges and edge-plane ratio of various TMDs at the atomically sharp limit. The process combines standard top-down lithography nanofabrication methods with anisotropic wet etching. This anisotropic etching process operates in a wide parameter range, from bulk crystals down to bilayers, and allows fabricating nanostructures with various sizes and shapes ranging from a single isolated hexagonal hole to densely packed arrays of holes with lateral edge-to-edge separations as narrow as ~3 nm. Therefore, a variety of unexplored TMD nanostructures and metamaterials, such as semiconducting-plane metallic-edge composites and ultrathin nanoribbons, can be fabricated in a highly controllable and scalable fashion. Notably, the method operates at ambient conditions and uses only abundant and cheap chemicals.

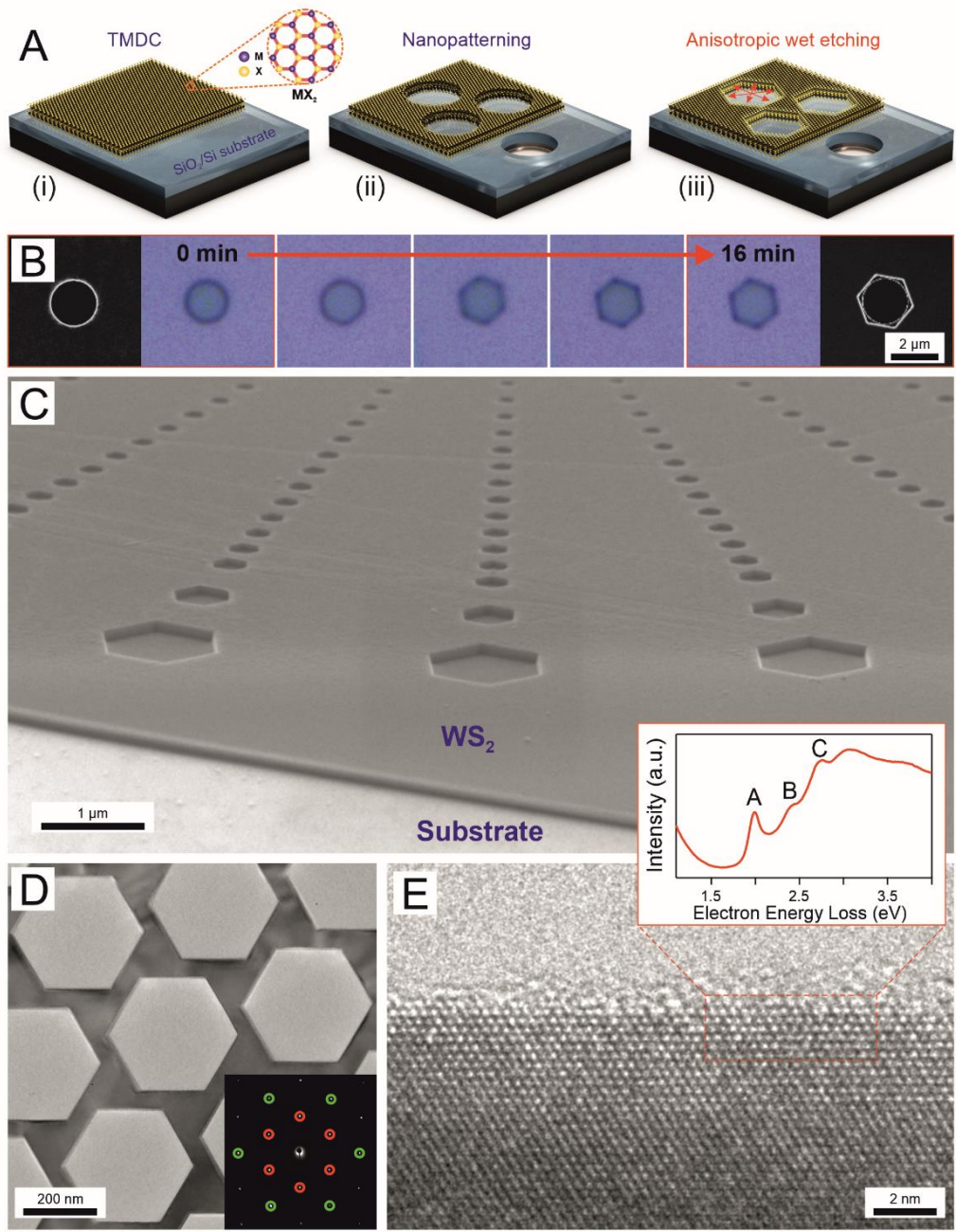

**Fig. 1. Etching hexagonal nanostructures in TMD materials.** (A) Schematic of the fabrication method: (i) Transfer of a mechanically-exfoliated TMD flake on a substrate, (ii) nanopatterning of the transferred flake using standard top-down lithography, and (iii) anisotropic wet etching of the pre-patterned flake. (B) Time evolution of the anisotropic etching process for a single hole in a $WS_2$ multilayer flake. Leftmost and rightmost panels show SEM images of the hole after 0 min and 16 min of anisotropic wet etching, respectively. (C) $70^0$ tilted SEM image of the etched $WS_2$ flake, revealing all of the etched hexagonal holes are oriented in same direction. (D) TEM image of an etched $WS_2$ hexagonal array (Inset shows a selected area electron diffraction pattern, revealing a single crystalline $WS_2$ diffraction pattern along a [0001] zone axis with an orientation that confirms the etched surfaces have zigzag character. The red and green circles identify the 1100 and 1120 families of diffraction spots, respectively). (E) HRTEM image of an etched surface, revealing its atomically sharp zigzag nature (Inset shows

an STEM EELS spectrum acquired with the electron probe positioned directly at an etched zigzag surface, revealing the presence of strong A-, B-, and C-exciton signals).

The three-stage fabrication process is summarized in Fig. 1A (more details are shown in Fig. S1). In stage (i), a TMD flake, such as for example tungsten disulfide ($WS_2$), of a certain thickness is isolated from an as-grown crystal onto a thermally oxidized silicon substrate using standard mechanical exfoliation (*25*). In stage (ii), the exfoliated flake is patterned using conventional top-down methods, such as a combination of electron beam lithography (EBL) and reactive ion etching (RIE), or by a focused ion beam (FIB). Such kind of patterning has been previously used to obtain various nanophotonic (*6*) and excitonic (*26*) structures. However, the nanostructures obtained in this way are not well-defined in terms of edges, which consist of an uncontrolled mix of zigzag, armchair and, possibly, amorphous configurations. In stage (iii), to achieve atomically sharp edges with well-defined terminations, the flake that was pre-patterned in stage (ii) is further exposed to an etchant solution. This exposure results in highly anisotropic wet etching of the pre-patterned circular holes in the flake, with etching occurring preferentially in-plane (Fig. 1A).

After a few minutes of etching, the pre-patterned circular holes transform into regular hexagonal holes (Fig. 1B). As seen from optical and scanning electron microscopy (SEM) images (Fig. 1B and Fig. S2), an initial single circular hole with a radius of ~1 µm is transformed into a perfect hexagonal hole within ~8 min. The final etching time can be accurately controlled by temperature, concentration and composition of the etching solution (Fig. S2). A $70^0$ tilted SEM image shows that the obtained hexagons have sharp sidewalls and are all oriented in the same way (Fig. 1C). This orientation follows the hexagonal crystallographic symmetry of $WS_2$ crystal along [0001], as we show below. The hexagonal shape also implies that the etching mechanism is self-limited by certain crystallographic planes, similar to the well-known anisotropic wet etching of single crystalline silicon (*27*).

To verify the sharpness of the obtained edges, we performed a transmission electron microscopy (TEM) study, which shows that the etched sidewalls are nearly atomically sharp with surface roughness variations of 1-2 atomic planes (Fig. 1D-E and Fig. S4). Selected area electron diffraction reveals the orientation of the etched hexagons are always synced to the TMD crystallography such that the etched surfaces are perpendicular to the <1100> direction (Fig. 1D). This confirms that the etched surfaces are exclusively zigzag-terminated. This observation agrees with earlier DFT results, which predict that zigzag edges are more stable than armchair ones (*14, 28*). Low-loss electron energy loss (EEL) spectra acquired from the etched surfaces show strong signals from A-, B-, and C-excitons (inset in Fig. 1E and Fig. S5), confirming that the $WS_2$ material is of high quality and the excitonic properties are undamaged by the etching process to the very edge of the material. Low-loss EELS spectra from the etched surfaces also reveal the presence of a signal at ~14 eV (Fig. S6) that is similar to a signal which has been previously associated with zigzag edges in $MoS_2$ (*29*). TEM images reveal that on top of the etched zigzag surfaces a thin layer (0-3 nm) of amorphous material occasionally appears (Fig. 1E and Fig. S4). This material is likely due to left-over residues originating either from the etching process or more likely from the scotch-tape and PDMS stamps used during the dry-transfer process for TEM sample preparation (see Methods). Additional data about the quality of the edges is given in the SI (Figs. S4-S8).

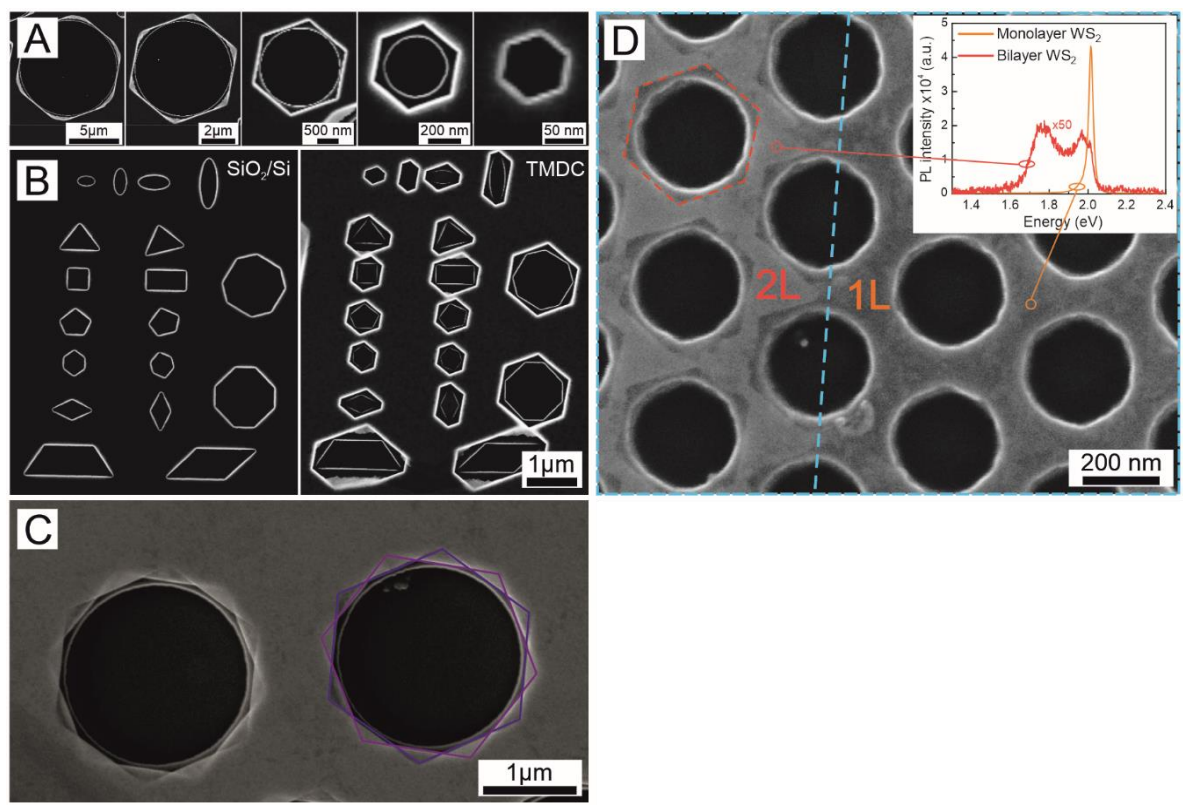

**Fig. 2. Controlling the morphology of a single etched hole.** (A) Effect of circular hole diameter: SEM images of etched holes in $WS_2$ with various sizes from ~50 nm to ~10 μm in diameter. (B) Effect of hole shape: SEM images of single holes with different initial shapes, such as pentagons, octagons, and ellipsoids, etched into both a bare $SiO_2$/Si substrate and a $WS_2$ flake, illustrating the comparison between initial and anisotropically-etched holes. (C) Effect of heterostructure: SEM image of etched hexagonal holes in a $WS_2$ multilayer heterostructure that has a $30^0$ rotation between the stacked flakes. (D) Effect of flake thickness: SEM image of hexagonal hole

arrays in bi- and monolayer WS$_2$. The inset shows the photoluminescence spectra of bi- (red curve) and monolayer (orange curve) WS$_2$.

To explore the morphological diversity and the scalability of the proposed etching method, we first investigated the behavior of single isolated nanoholes. The hole size and shape were initially of concern (Fig. 2A-B respectively). Fig. 2A and Fig. S9 show that it is possible to etch a WS$_2$ multilayer to produce hexagonal etched holes with diameters varying in a broad range – from 50 nm to 10,000 nm (tested here, however, there seem to be no limitation on the hole size from above). It is also important to mention that small circular holes rapidly transform into perfect hexagonal holes, whose subsequent etching is almost entirely stopped. Larger holes, however, converge to perfect hexagons at a much slower pace. Thus, if the etching is stopped before the perfect hexagonal shape is reached, the resulting hexagonal holes appear incomplete.

In addition to the effect of the hole size, we have examined the effect of hole shapes (Fig. 2B). A variety of non-circular shapes, including ellipses, triangles, squares, pentagons and trapezoids were tested and all them were converted into regular or irregular hexagonal shapes, irrespectively of the shape and size of the original hole (Fig. 2B). Among other things, this suggests that random defects on a TMD surface can be anisotropically etched and converted into perfect zigzag-terminated features using this anisotropic etching method. Additional data about the quality of single holes, more complicated structures such as bull's eye nanostructures, as well as tests performed on multiple semiconductor group VI TMD materials is given in the SI (Figs. S9-S11).

We further investigated a single hole in a TMD heterostructure (Fig. 2C), which as it turns out can also be etched into a hexagon, despite the fact that the crystallographic axes in the heterostructure are rotated with respect to each other. Etching a two flake heterostructure where the top flake (WS$_2$) was rotated by ~$30^0$ with respect to the bottom flake (another WS$_2$) resulted in hexagons rotated by ~$30^0$ (Fig. 2C and Fig. S12). This confirms the etching is anisotropic even for heterostructures and suggests that this method can be used to reveal the crystallographic orientation of different flakes with respect to each other.

To verify the influence of the flake thickness on the etching process, we have investigated several different thicknesses, ranging from a few hundred nanometers down to a monolayer. Monolayer TMDs have attracted substantial attention in recent years due to their direct bandgap and interesting excitonic physics (*1, 2*). As shown in Fig. 2D and Fig. S13, a monolayer of WS$_2$ withstands the etching process, as evident from a bright photoluminescence (PL) signal. However, holes in a monolayer cannot be directly patterned into hexagonal shapes, instead they are transformed into irregular triangular shapes (Fig. S13B-E). In contrast, an adjacent bilayer $WS_2$ is etched into hexagonal shapes, as shown in the SEM image in Fig. 2D (the bilayer nature is confirmed by the PL signal). This result has three important implications. First, it confirms that the etching mechanism has an anisotropic nature and does not etch along the basal plane [0001], since the bilayer remains the bilayer (and does not get etched into a monolayer or further) throughout the entire etching process. Second, it signals that in addition to different optical properties, monolayers might also have different chemical properties in comparison to multilayers. Despite the impossibility of direct hexagonal etching of monolayers, such patterned monolayers can be fabricated by exfoliating etched multilayers using a variety of well-established techniques (*25*). Third, it highlights the uniqueness of the proposed etching method as it operates exclusively with multilayers, starting from bilayers and up to hundreds of nanometers (thicker samples were not tested in this study). This implies that the process of hexagonal etching inherently requires a multilayer nature of the TMD flake.

To explore more complicated and extended structures, as well as to test the scalability of our approach, we fabricated regular arrays of hexagonal nanoholes (Fig. 3). Our results show that the desired geometry can be accurately and deterministically controlled over the entire flake (up to several hundreds of microns and is limited by the flake size), thus forming an extended TMD metamaterial with atomic precision in the structure. Moreover, the pattern quality is limited not by the size of the flake itself, but rather by the methods used to pre-pattern the sample in stage (ii), that is, by EBL and RIE (Fig. 3 and Fig. S14-17). Various periodic patterns, such as honeycombs, vortexes, and bow-tie arrays, can be fabricated by controlling the angle between the array and the crystallographic axis of the $WS_2$ material (Fig. 3A and Fig. S16). The orange-dashed and blue hexagons in Fig. 3A outline the orientations of hexagonal lattice and individual hexagonal holes, respectively. The precise control over the parameters of the periodic lattice with respect to the given orientation of individual hexagonal holes (which follows the crystallographic orientation of the $WS_2$ flake) enables fabrication of a rich variety of nanostructured arrays (Fig. 3A). The side walls in these arrays are as sharp as the single nanohole structures shown previously, suggesting the method is scalable. Furthermore, because the etching time and the initial circular hole size and location can be accurately controlled, a variety of complex TMD nanostructures with preferential zigzag edges can be readily fabricated over large areas (Fig. 3B and Fig. S16). Importantly, many of these nanostructured arrays are challenging to obtain using alternative methods, whereas the method described here combines a high level of versatility, precision, and control with scalability over macroscopic dimensions.

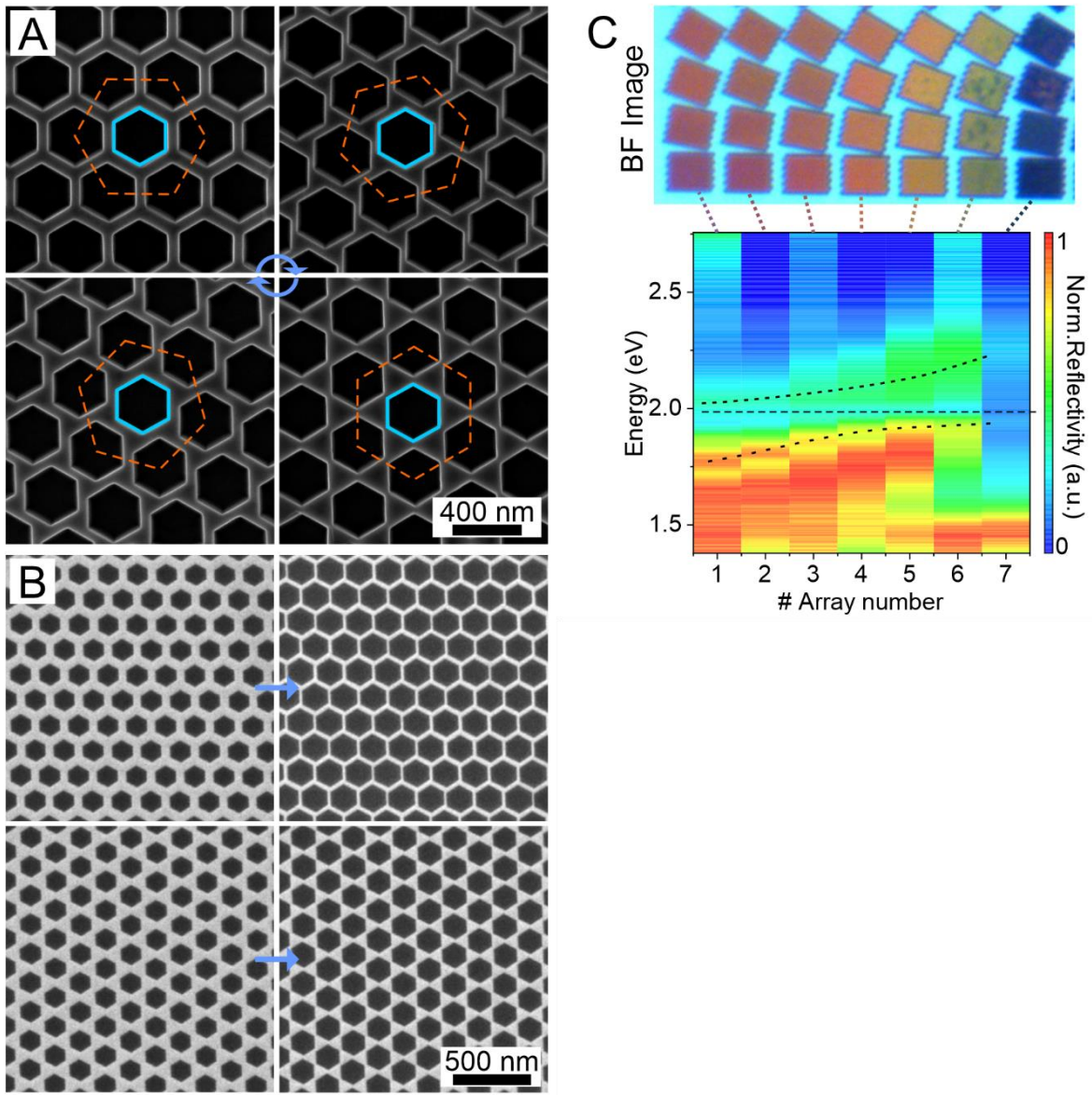

**Fig. 3. A variety of etched hexagonal hole arrays and their optical properties.** (A) SEM images of hexagonal hole arrays with various rotational angles of a $15^0$ step with respect to the orientation of individual hexagonal holes in a WS$_2$ flake. The orange-dashed and blue hexagons identify the orientations of hexagonal lattice arrays and individual hexagonal holes, respectively. (B) SEM images of honeycomb and bow-tie hexagonal hole arrays with fixed lattice periodicity but different hole sizes (bigger holes are shown to the right). (C) Optical microscope image and reflectivity spectra of hexagonal arrays with various hole sizes and lattice orientations in ~70 nm thick WS$_2$. The bottom row of the optical image consists of honeycomb arrays with variable hole sizes ranging from 240 nm to 300 nm (left to right) and a fixed lattice pitch of 400 nm. The corresponding reflectivity spectra are shown in the color plot below. The top rows of the optical image consist of hexagonal hole arrays that are rotated by $10^0$ steps (for every next row) with respect to the bottom row.

Optical images of the obtained hexagonal lattices in ~70 nm thick $WS_2$ (Fig. 3C) reveal bright structural colors, which are dependent on hole size and arrangement. This suggests that nanopatterned $WS_2$ arrays act as photonic crystals, making them interesting for structural color applications and tailored light-matter interactions. Importantly, many TMD materials such as $MoS_2$ and $WS_2$ have exceptionally high values of the in-plane refractive index $n>4$ (*6, 30, 31*), which, together with the etching method discussed here, enrich the TMD nanophotonics toolbox. The bottom row of the optical image in Fig. 3C represents honeycomb arrays composed of various initial hole sizes ranging from 240 nm to 300 nm with a step of 10 nm and a fixed lattice pitch of 400 nm. Their corresponding reflectivity spectra are shown in the color plot at the bottom of Fig. 3C. These reflection spectra reveal interesting resonant features, which linearly disperse with the hole size in the near-infra red region (<1.9 eV). The mode shows additional avoided-crossing around the $WS_2$ A-exciton (~2 eV), which can be potentially attributed to hybridization between optical modes of the array and $WS_2$ exciton, similar to a previous study on single $WS_2$ nanodisks (*6*). Additionally, array #7 exhibits a weak reflectivity (~15%) over the entire visible spectrum, suggesting that it is highly absorptive and thus could be useful for light harvesting applications. The top rows of optical images in Fig. 3C correspond to arrays composed of hexagonal holes that are rotated in $10^0$ steps with respect to the bottom row. These structures thus correspond to a continuous transition between honeycomb to bow-tie shaped arrays.

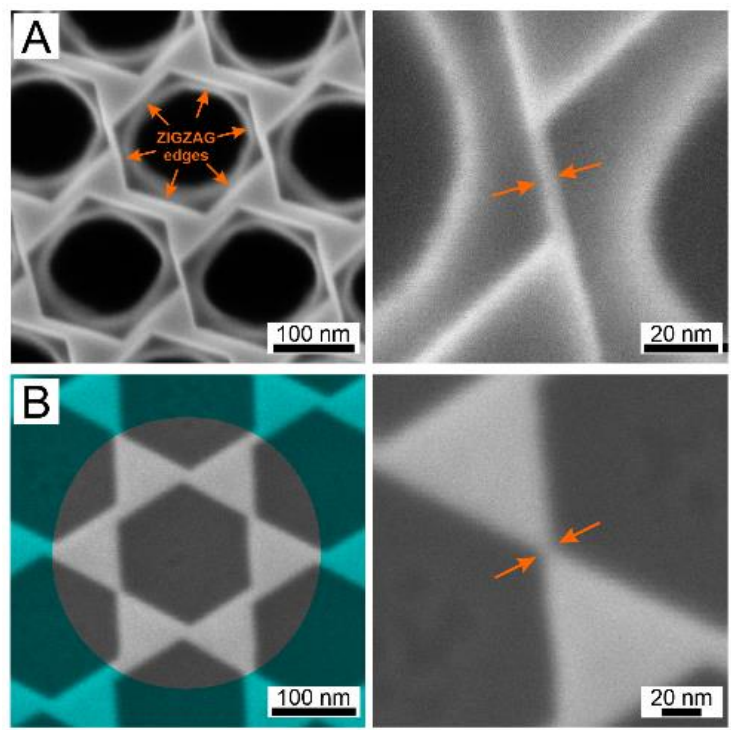

**Fig. 4. Fabrication of a few nanometer narrow nanoribbons and nanojunctions.** (A) SEM images of a (left-panel) hexagonal hole array in $WS_2$ and a (right-panel) thin (~3 nm) nanoribbon that consists of $WS_2$ sandwiched between two zigzag-terminated edges. (B) SEM images of a (left-panel) hexagonal hole array in $WS_2$ resulting in bow-tie nanostructures and a (right-panel) thin $WS_2$ nanojunction structure.

A combination of accurate control over the etching rate and the atomic sharpness of the etched zigzag edges enables fabrication of various narrow nanostructures (Fig. 4). The top row in Fig. 4 shows nanoribbons with remarkably sharp vertical walls in densely packed arrays. It has been theoretically predicted that ultrathin TMD nanoribbons have drastically different optical, excitonic, and electronic properties compared to conventional TMD materials (*13, 28*). However, experimental evidence for this is lacking because of the difficulty in fabricating these nanostructures. Furthermore, the bowtie array etched down to a sharp vertical wall limit results in narrow nanojunctions (Fig. 4B). The obtained nanoribbons and nanojunctions can have edge-to-edge distance of about ~3 nm. Importantly, even smaller nanoribbons and nanojunctions could be potentially possible since the currently obtained size features are limited by the quality of the initial patterns created by the conventional top-down nanofabrication methods. The sizes of these nanoribbons and nanojunctions are compact enough to approach the size of the exciton Bohr radius in $WS_2$ multilayers, which could lead to additional quantum confinement effects. Furthermore, these structures have a high surface-to-volume ratio, which could be useful for catalytic (*7-9*) and nonlinear optics (*17*) applications.

In summary, we have demonstrated a new anisotropic wet etching method that enables fabrication of complex hexagonal nanostructures with atomically sharp zigzag edges in various TMD materials ($WS_2$, $WSe_2$, $MoS_2$, and $MoSe_2$ are tested here) using only abundant and cheap chemicals at ambient conditions. Our method allows for fabricating novel single nanostructures, complex nanostructured arrays, and vertically stacked ultrathin TMD nanoribbons, which to the best of our knowledge is not possible using other conventional techniques. We unite all these structures under a common umbrella of TMD metamaterials.

We envision that the combination of high surface-to-volume ratio and exclusively zigzag edges, which arise in these TMD metamaterials, could be useful for multiple purposes, including TMD-based catalysis (*7, 8*) and sensing (*32*) applications. Furthermore, there is a potential for other areas and applications, such as photodetectors (*33*), quantum transport (*4*), nonlinear optics (*17*), nanophotonics (*6*), and optomechanics (*34*) to benefit from the method.

Finally, we speculate that it may be potentially possible to generalize this method to other 2D material classes, such as graphene, hexagonal boron nitride, Mxenes, and others, because the anisotropic etching mechanism is likely related to different relative stabilities of various edges and planes (zigzag, armchair, basal), which is a generic property for any vdW material with hexagonal crystallographic symmetry.